\documentclass[11pt]{article}
\usepackage[colorlinks]{hyperref}
\usepackage[authoryear]{natbib}
\usepackage{graphicx}
\usepackage{amsmath, amsthm, amssymb}
\usepackage{xcolor}
\usepackage{hyperref}
\usepackage[top=1in, bottom=1in, left=1in, right=1in]{geometry}

\newenvironment{enum}{
\begin{enumerate}
  \setlength{\itemsep}{1pt}
  \setlength{\parskip}{0pt}
  \setlength{\parsep}{0pt}
}{\end{enumerate}}

\begin{document}

\begin{center}
\textbf{Matrix eQTL: Ultra fast eQTL analysis via large matrix operations}

Andrey A. Shabalin

Department of Biostatistics, University of North Carolina at Chapel Hill
\end{center}

\textbf{Abstract.} 

\vskip 0.1in

\textbf{Motivation:}
Expression quantitative trait loci (eQTL) mapping aims to determine genomic regions that regulate gene transcription. 
Expression QTL is used to study the regulatory structure of normal tissues and to search for genetic factors in complex diseases such as cancer, diabetes, and cystic fibrosis.

A modern eQTL dataset contains millions of SNPs and thousands of transcripts measured for hundreds of samples. This makes the analysis computationally complex as it involves independent testing for association for every transcript-SNP pair. 
The heavy computational burden makes eQTL analysis less popular, often forces analysts to restrict their attention to just a subset of transcripts and SNPs. As larger genotype and gene expression datasets become available, the demand for fast tools for eQTL analysis increases.
%The analysts often have to run the analysis only once, 
%The computational burden is even greater when non-linear modeling is performed or if the significance of the findings is assessed using permutations. As larger genotype and gene expression datasets become available, there is an increased demand for fast tools for eQTL analysis.
%the need for fast tools for eQTL analysis is growing stronger.

\textbf{Solution:} 
%In this paper 
We present a new method for fast eQTL analysis via linear models, called Matrix eQTL. 
Matrix eQTL can model and test for association using both linear regression and ANOVA models.
The models can include covariates to account for such factors as population structure, gender, and clinical variables.
It also supports testing of heteroscedastic models and models with correlated errors.
In our experiment on large datasets Matrix eQTL was thousands of times faster than the existing popular software for QTL/eQTL analysis. 
%errors, and for permutation based testing.  %These features will be added in the next version of the software.

Matrix eQTL is implemented as both Matlab and R packages and thus can easily be run on Windows, Mac OS, and Linux systems. The software is freely available at the following address.

\textbf{Website:} \url{http://www.bios.unc.edu/research/genomic_software/Matrix_eQTL}

\vskip 0.1in

\textbf{Author Summary.}

\vskip 0.1in

Expression quantitative trait loci (eQTL) analysis is a part of genetical genomics linking variations in gene expression levels to the differences in genotype. It is important for understanding of the regulatory functioning in both normal and diseased tissues. Modern eQTL studies measure gene expression over tens of thousands of transcripts and variations of genotype (single nucleotide polymorphisms, SNPs) over millions of markers. This draws the analysis extremely computationally intensive as it involves independent testing for association for every transcript-SNP pair. Computational burden makes eQTL analysis less popular, often forces analysts to restrict their attention to a small subset of genes-SNP pairs. 

We present a new fast tool for eQTL analysis called Matrix eQTL. It is designed to handle large datasets and is hundreds to thousands times faster that the existing tools for eQTL analysis. 
%Matrix eQTL brings multiple benefits to the analysts. 
Using Matrix eQTL analysts can fit and compare multiple models in a matter of minutes or hours, even for large datasets. Matrix eQTL makes it fast and simple to test different preprocessing and quality control procedures. One can also use Matrix eQTL to determine the right significance thresholds via permutation testing. Matrix eQTL will greatly improve the ability of analysts to find true eQTLs by fitting the right model and determining the correct significance threshold.

\vskip 0.1in

\vskip 0.1in

\textbf{Introduction.}

\vskip 0.1in

%\textbf{eQTL.} 
The goal of eQTL analysis is to identify genomic locations where the genotype is significantly associated with expression of known genes. These associations can help reveal biochemical process underlying living systems, discover the genetic factors causing certain diseases and determine pathways that are affected by them. Expression QTL analysis is used to determine hotspots (DNA regions affecting multiple transcripts, \cite{breitling2008genetical}), construct causal networks, discover subclasses of clinical phenotypes and select genes for clinical trials \citep[see reviews of][]{gilad2008revealing,zhang2010qtl, kendziorski2006review}.

%\textbf{Approaches.} 
There are various approaches to eQTL analysis. Some methods test for transcript-SNP associations independently, while others aim to find multiple SNPs that can jointly explain variations in expression of a gene. There is a wide range of models used to fit the expression values. 
In this paper our goal is fast analysis of large datasets and thus we focus on independent testing of each transcript-SNP pair using linear regression.
More elaborate approaches include non-linear modeling, such as generalized linear models, Bayesian models, and models accounting for pedigree.

%\textbf{Complexity.} 
Expression QTL analysis is known to be computationally intensive. 
The issue is most pronounced for modern eQTL dataset, which often contain millions of SNPs and thousands of transcripts measured for hundreds of samples.  
For such data, the complete eQTL analysis may involve over ten billion tests.
%Pairwise linear regression analysis with existing tools takes a lot of time for such data (see Results Section). 
The following three studies indicate that non-linear methods can be prohibitively slow for large datasets. 
%Authors of the following three papers reported timing for their analyses.
First, \cite{degnan2008genomics} were applying Family Based Association Tests (FBAT) to a dataset with 142 samples. They tested only 40,227 transcript-SNP pairs and report the total computation time to be `under 24 hours on 20 processors in parallel on a Linux cluster'. Second, \cite{ghazalpour2008high} were running Efficient Mixed-Model Association (EMMA), which is claimed to be computationally efficient. They analyzed a dataset with 110 samples, 1,813 SNPs, and 10,013 transcripts. They tested all transcript-SNP pairs and report the computation time of `a few hours using a cluster of 50 processors'.
Third, \cite{listgarten2010correction} tested their method on a dataset containing 40,639 probes in the expression data and 48,186 SNPs for 188 samples. They report that estimation took 10 hours when parallelized across 1,100 processors (this is more than a processor-year). Note that the dimensions of a modern eQTL dataset can greatly exceed those in the examples above.

%\textbf{Intro to Matrix eQTL.}
For a large dataset, eQTL analysis using existing tools requires multiple processor-weeks, and thus often has to be performed on a computing cluster. 
In this paper we present a new ultra fast method, called Matrix eQTL, for eQTL analysis using linear models.
%Matrix eQTL is optimized for analysis of large datasets. 
Matrix eQTL allows to analyze large data on a single desktop machine in the time existing tools would run on a large computing cluster.
In our tests on a cystic fibrosis dataset with 840 samples Matrix eQTL strongly outperformed existing QTL/eQTL tools.

Matrix eQTL performs separate testing for each transcript-SNP pair. The association between each SNP and each transcript is tested using either least squares regression or ANOVA model. 
%Matrix eQTL can use linear regression and ANOVA models to test for association. 
Both linear regression and ANOVA models can include extra additive covariates to account for such factors as population structure, gender, and clinical variables.
Matrix eQTL also supports such extensions of linear regression model as weighted least squares, model with correlated errors, and mixed effects model with known variance parameters. %These options will be implemented in the next version of the software.

Matrix eQTL uses a special algorithm and data preprocessing allowing testing for the association between the genotype and expression without estimation of all the model parameters. The most computationally intensive part of the algorithm is formulated in terms of operations with large matrices. This allowed us to implement the fast algorithm using high-level programming languages, R and Matlab, relying on their efficient implementation of matrix operations. The performance of Matrix eQTL depends on the efficiency of matrix multiplication routine. More information about matrix multiplication and its implementations in R and Matlab is provided in the Appendix.

Matrix eQTL is cross platform and can be run on any platform for which Matlab or R is available, namely Linux, Mac OS X, and Windows.
The Matlab and R implementations of Matrix eQTL are independent and have equal functionality. The performance of the implementations may differ depending on the version of Matlab or R used to run the code.

\vskip 0.1in

\textbf{Results.}

\vskip 0.1in

\textbf{Data.} 
To assess the performance of Matrix eQTL and compare it to other eQTL tools we used genotype and gene expression data from 840 patients with cystic fibrosis \cite{wright2011genome}.% \cite{bartlett2009genetic}.
%The dataset contains genotype and gene expression information for 840 common samples. 
The genotype information was obtained for 573,337 markers and the gene expression was measured for 22,011 transcripts.
The missing values were imputed by average values of the variable across samples.

\textbf{Performance.}
We compare performance of Matrix eQTL with that of five programs for QTL and eQTL analysis: Plink \citep{purcell2007plink}, Merlin \citep{abecasis2001merlin}, R/qtl \citep{broman2003r}, eMap \citep{sun2009eqtl}, and FastMap \cite{gatti2009fastmap,gatti2011fastmap}.
Plink is a command line toolset for whole genome association analysis. It is written in C/C++ and as a general purpose tool is not optimized for eQTL analysis. 
Merlin is a command line tool for fast pedigree analysis written in C/C++. It is designed for analysis in pedigree and may not show the best performance for unrelated samples.
R/qtl and eMap are R packages for QTL/eQTL analysis, part of eMap is coded in C and requires GSL library (GNU Scientific Library).
FastMap is a user friendly tool for fast association mapping written in Java. It uses discrete nature of genotype data to speed up calculations, but it can not handle covariates. FastMap is optimized for permutation based testing. Note that FastMap is the only tool out of the five that has a graphical user interface.

All methods except R/qtl were set to estimate the simple linear model for the relationship between gene expression and genotype. R/qtl does does support the simple linear model and was set to estimate the ANOVA model (Haley-Knott method).
First, we ran all methods on a subset of the cystic fibrosis data, containing 2,000 random genes and SNPs. 
To reduce the output we set, where possible, the p-value threshold at $10^{-5}$ level. The technical specifications of the machine used for testing are provided in the Appendix.
Table \ref{tabperf} shows that the first four existing methods performed the analysis in more than eight minutes and FastMap finished in about one minute, while both versions of Matrix eQTL (Matlab and R) finished in less than half of a second. 

The time required for the analysis of the complete datasets is presented in the right column of the table. For the five existing methods the time is estimated under the assumption of linearity with respect to the number of transcripts and the number of SNPs. Both implementations of matrix eQTL were applied to the complete dataset to obtain the precise timing. % Note that timings of Matrix eQTL for the 2k $\times$ 2k subset were both under half a second, and thus not very precise.
\begin{table}[[h]
\begin{center}
\begin{tabular}{lrl} 
					& \multicolumn{1}{l}{2k Genes} & Complete \\
Method & \multicolumn{1}{l}{2k SNPs} & dataset \\ \hline
Plink & 2678 sec & 97.8 days \\ %\hline
Merlin & 564 sec & 20.6 days \\
R/qtl & 596 sec & 21.2 days \\ %\hline
eMap & 492 sec & 18.0 days \\ 
%FastMap & \phantom{0}51 sec & \phantom{0}1.9 days \\
FastMap & \phantom{0}61 sec & \phantom{0}2.2 days \\
\hline
Matrix eQTL (Matlab) & 0.23 sec & 11.5 minutes \\ %\hline
Matrix eQTL (Rev R)  & 0.47 sec & 13.4 minutes \\ %\hline
\end{tabular}
\caption{Performance of various eQTL software on the Cystic Fibrosis dataset. The time for first 4 methods is projected from a random subset of 2,000 genes and 2,000 SNPs. %The timing for the subset is presented in Supplementary Materials.
\label{tabperf}}
\end{center}
\end{table}

Matrix eQTL can perform analyses that previously required days or weeks in just minutes.  QTL modelers that have only a handful of phenotypes understand the importance of testing several different models with covariates and interaction terms.  Until now, this has not been computationally feasible with eQTL analyses due to the computational burden.  Matrix eQTL allows analysts to fit a variety of models with different mixes of covariates in less than an hour and compare the results.  Likewise, it allows analysts to compare different preprocessing and quality control procedures. Furthermore, once an appropriate model has been selected, Matrix eQTL can be used to determine permutation based significance thresholds in less time than most packages take to generate nominal p-values. Matrix eQTL will greatly improve the ability of analysts to find true eQTLs by fitting the right model and determining the correct significance threshold.

\vskip 0.1in

\textbf{Methods.}

\vskip 0.1in

We describe the algorithm of Matrix eQTL in steps. First, we detail the algorithm for the simple linear regression. Then we show how the algorithm is extended to handle the ANOVA model and covariates. Finally we show how the algorithm is extended to test heteroskedastic models and models with correlated errors.
\vskip 0.1in
\textbf{Simple linear regression.} The simple linear regression is probably the commonly used model for eQTL analysis. For each transcript-SNP pair, the association between gene expression $g$ and genotype $s$ is assumed to be linear, with genotype encoded as $0,1$, or $2$ according to the frequency of the minor allele.
\begin{equation}
	g = \alpha + \beta s + \epsilon, \quad \mbox{ where } \ \epsilon \sim i.i.d.\,N(0,\sigma^2)
\label{modelb}
\end{equation}

The conventional algorithm for the analysis of the simple linear regression includes estimation or calculation of a number of parameters: the sample means $\bar g$ and $\bar s$, the slope coefficient $\hat \beta$, the intercept $\hat \alpha$, the residuals $e_i$, the total sum of squares $SST$, and the residual sum of squares $SSE$. This is followed by the estimation of a test statistic, which can be t-statistic, F-test, or the likelihood ratio test. Finally, the p-value is calculated for the test statistic; this step can also be computationally intensive as it involves calculation of incomplete beta or gamma functions.

The goal of Matrix eQTL is to find all transcript-SNP pairs with association significant at a given level. This allows us to skip estimation of unnecessary parameters and focus on the most efficient calculation of a test statistic.

To save time, Matrix eQTL does not calculate p-value for every transcript-SNP pair. 
Instead, for the test statistic of choice, it finds the threshold, above which the test statistic is significant at the required significance level. The test statistics for every transcript-SNP pair are then compared to the threshold, and the p-values are calculated only for those above the threshold.

The choice of test statistic is important performance of the method. It is natural to choose the test statistic that can be calculated faster, among statistics of equal power. Observe that for the simple linear regression (\ref{modelb}), the common statistics, such as $t$, $F$, $R^2$, and $LR$, are equivalent and can be expressed as functions of the correlation $r = \mbox{cor}(g,s)$
\[
 t = \sqrt{n-2}\frac{r}{\sqrt{1-r^2}},
\qquad
 F = t^2 = (n-2)\frac{r^2}{1-r^2},
\qquad
 R^2 = r^2,
\qquad
 LR = - \log(1-r^2).
\]

Thus we choose the absolute value of the correlation $|r|$ as the test statistic for the simple linear regression and threshold it to find significant transcript-SNP associations. Note that correlation does not change if we standardize the genotype and gene expression variables such that 
\[
 \sum g_i = 0, 
\qquad 
 \sum g_i^2 = 1,
\qquad
 \sum s_i = 0,
\qquad
 \sum s_i^2 = 1.
\]

\begin{figure}[h]
\begin{center}
\includegraphics[height=2.5in]{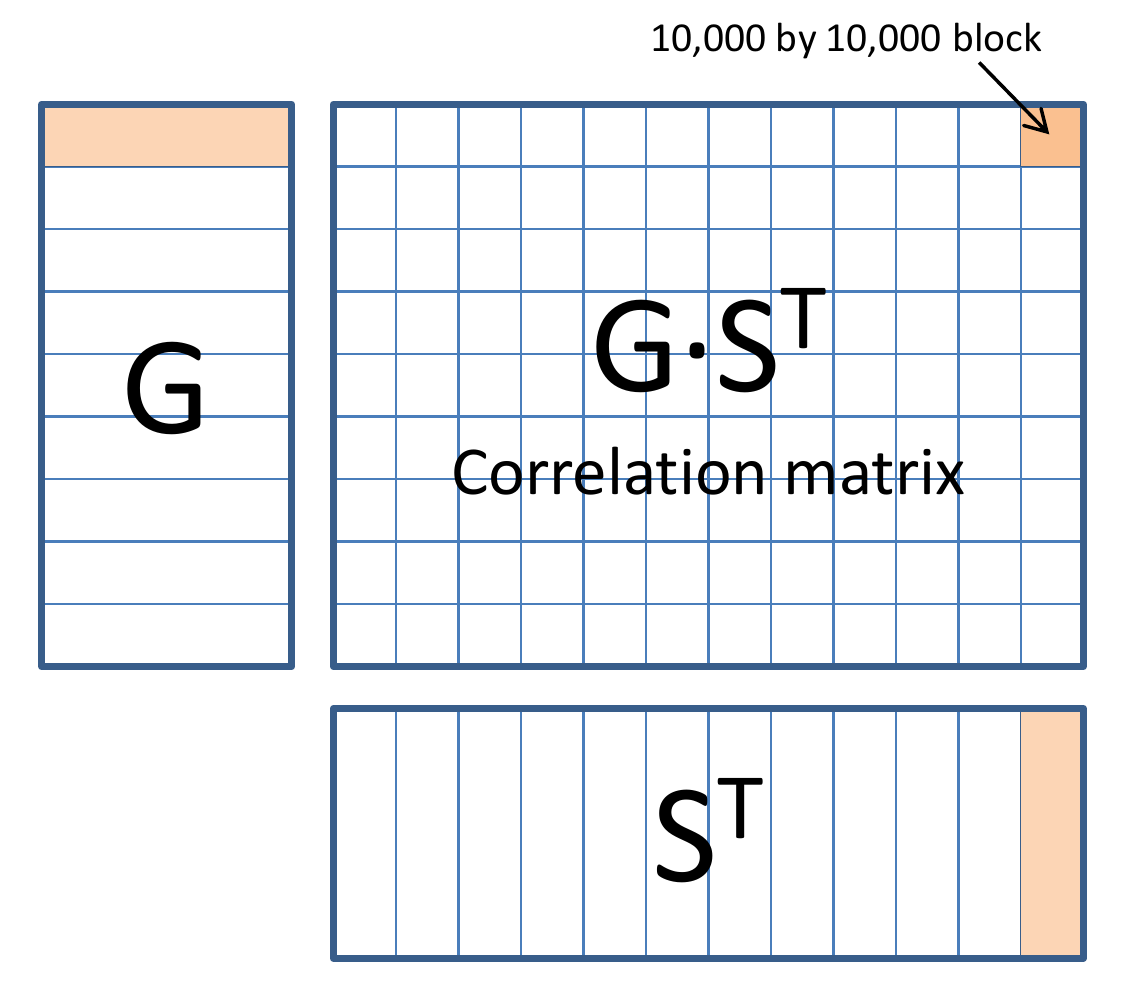}
\caption{Matrix of correlations can be calculated using matrix multiplication. Due to the huge number of tests the analysis performed in 10,000x10,000 blocks.
\label{figCor}}
\end{center}
\end{figure}

The standardization does not add complexity to the calculations as it has to be performed only once for each transcript and once for each SNP, and it greatly simplifies the calculation of the correlation
\[
r = \mbox{cor} (s,g) = \frac{\sum(s_i - \bar s)(g_i - \bar g)}{\sqrt{\sum(s_i - \bar s)^2\sum(g_i - \bar g)^2}} = \sum s_i g_i  = \langle s,g \rangle
,
\]
where $\langle s,g \rangle$ denotes the inner product between vectors $s$ and $g$.
Now, let $S$ be the genotype matrix, with each row containing measurements for a single SNP and each column containing measurements for a single sample. Let $G$ be the gene expression matrix, with each row containing measurements for a single transcript and each column containing measurements for a single sample. Let the columns (samples) of matrix $S$ match those of matrix $G$. Then the matrix of all gene-SNP correlations can be calculated in just one matrix multiplication $GS^T$ as illustrated in Figure \ref{figCor}. 

However, the number of tests in a modern eQTL study may exceed tens of billions and such correlation matrix would require hundreds of gigabytes of RAM. To avoid excess memory requirements we slice the data matrices in blocks of up to 10,000 variables and perform the analysis separately for each pair of blocks. Figure \ref{figCor} illustrates the slicing and calculation of the correlation matrix.

\vskip 0.1in

%\textbf{Algorithm.} 
The algorithm of Matrix eQTL for the simple linear regression is as follows:
\begin{enum}
	\item Split input matrices into blocks of up to 10,000 variables.
	\item Standardize variables of both gene expression and genotype matrices.
	\item For each pair of blocks:
	\begin{enum}
		\item Calculate the corresponding block of the correlation matrix in one matrix operation
		\item Find absolute correlations exceeding predefined threshold
		\item For those correlations, calculate test statistic, p-value, and other variables of interest and report them
	\end{enum}
\end{enum}

\vskip 0.1in

\textbf{Model with covariates.} 
It is common to include covariates in the eQTL model to account for such effects as population stratification, gender, age, and  clinical variables. For simplicity, let's consider the model with one extra covariate $x$,
\begin{equation}
	g = \alpha + \gamma x + \beta s + \epsilon.
\label{modelcov}
\end{equation}
As for the simple linear regression, various test statistics testing for the significance of $s$, namely LR, F-test, and t-statistic, are equivalent. The testing can be reduced to the simple linear regression model by orthogonalization of $g$ and $s$ with respect to $x$.  The algorithm for the analysis is then:

\begin{enum}
	\item Center variables $g$, $x$, and $s$ to remove $\alpha$ from the model.
	\item Orthogonalize $g$ and $s$ with respect to $x$: 
$\ \tilde g = g - \langle g,x\rangle x, 
\ \tilde s = s - \langle s,x\rangle x.
$
	\item Perform the analysis for the simple linear regression
$
	\tilde g = \beta \tilde s + e
$
	using one less degree of freedom for the test statistic to account for the removed covariate.
\end{enum}

\textbf{ANOVA model.} Another common extension of the simple linear regression for eQTL analysis is to treat each genotype variable as categorical and model its effect on gene expression with ANOVA model. ANOVA model can be viewed as a linear regression with each SNP represented with two dummy variables: $s_1 = I(s=1)$ and $s_2 = I(s=2)$.
\begin{equation}
	g = \alpha + \beta_1 s_1 + \beta_2 s_2 + \epsilon 
\label{modelf}
\end{equation}
F-test or LR statistic are equivalent statistics for testing joint significance of $s_1$ and $s_2$. Both F and LR are monotone functions of $R^2$ for this model, thus we can use $R^2$ as the test statistic. The test statistic can be calculated efficiently if we orthogonalize the regressors. The algorithm for the analysis is then:
\begin{enum}
	\item Center variables $g$, $s_1$, and $s_2$ to remove $\alpha$ from the model.
	\item Orthogonalize $s_2$ with respect to $s_1$ for every marker, $\tilde s_2 = s_2 - \langle s_2,s_1\rangle s_1$, and standardize it.
	\item Use test statistic: $R^2 =  \langle g,s \rangle^2 + \langle g,\tilde s_2\rangle^2$
	\item The threshold for $R^2$ and p-values can be derived from the formula for F-test $F = \frac{(n-3)R^2}{2(1-R^2)}$.
\end{enum}
The same algorithm can be used to estimate the model with two marker-by-marker variables, such as genotype and copy number variations.

Another test statistics can be used to test for the significance of just one variable ($s_2$) accounting for the effect of another marker-by-marker variable ($s_1$). In this case the test statistic would be
\[
F =  \frac{n-3}{2} \cdot \frac{\langle g,\tilde s_2\rangle^2}{1 - \langle g,s_1\rangle^2 - \langle g,\tilde s_2\rangle^2}
.
\] 
This approach can easily be generalized for testing for joint significance of any subset of regressors.

\vskip 0.1in

\textbf{Heteroskedastic models and models with correlated errors.} The previously described models assume the noise to be independent and identically distributed across samples. However, the errors can be heteroskedastic if the quality of the measurements differs across samples. Also, the errors may be correlated if the samples come from a pedigree. To account for both possibilities, consider the model with non i.i.d. errors:
\begin{equation}
	g = \alpha + \beta s + u, \quad \mbox{ where } U \sim N(0,\sigma^2 K) 
\label{modelk}
\end{equation}
and $K$ is a known non-singular covariance matrix. To apply the previously described methods to this problem we transform the input variables to make the errors independent and identically distributed:
\[
	\tilde g = K^{-1/2} g,
\quad
	\tilde s = K^{-1/2} s,
\quad
	\tilde q = K^{-1/2} 1_n,
\]
where $1_n$ is a vector of ones. The new model equation is homoskedastic, has independent errors, but does not include a constant.
\begin{equation}
	\tilde g = \tilde q + \beta \tilde s + e, \quad \mbox{ where } e \sim i.i.d.\ N(0,\sigma^2) 
\label{modelk2}
\end{equation}
The model is tested using the algorithm for the linear model with covariates with step 1 (centering) omitted.
% The current version of Matrix eQTL allows testing using linear models with i.i.d. errors only. The heteroskedastic model and the model with correlated errors will be supported in the next version of Matrix eQTL.

\vskip 0.1in

\textbf{Grant Support}

\vskip 0.1in

This work was supported, in part, by funding from the National Institutes of Health (R01-MH090936 and R01-ES015241), US Environmental Protection Agency (STAR RD83382501 and RD83272001), National Cancer Institute (R01-CA138255), National Institute of Mental Health (R01-MH090936), and the Gillings Innovation Laboratory in Statistical Genomics.
The funders had no role in study design, data collection and analysis, decision to publish, or preparation of the manuscript.
\vskip 0.1in

\vskip 0.1in

\textbf{Conclusion}

\vskip 0.1in

In this paper we presented a new ultra fast method for eQTL analysis, called Matrix eQTL. The method can test for transcript-SNP associations using linear regression and ANOVA models with covariates. Is also supports models with heteroskedastic and correlated errors to account for sample quality and population structure. We tested Matrix eQTL and compared it to five existing eQTL methods. Matrix eQTL was hundreds to thousands of times faster than the other eQTL methods. 
Matrix eQTL allows one to perform eQTL analysis of large datasets on a single desktop machine in the time other methods would run on a large computing cluster. Matrix eQTL is implemented in Matlab and R, and the code is publicly available.

Fast performance of Matrix eQTL open new possibilities for analysts. 
Matrix eQTL does not require a computing cluster. Using Matrix eQTL analysts can fit and compare multiple models in a matter of minutes or hours, even for large datasets. Matrix eQTL simplifies testing of different preprocessing and quality control procedures. Matrix eQTL can also be to determine significance thresholds via permutation testing. Matrix eQTL will greatly improve the ability of analysts to find true eQTLs by fitting the right model and determining the correct significance threshold.

For the future versions of Matrix eQTL we plan to consider several extensions and modification. First, we consider performing calculation on a GPU (graphics processing unit), instead of the CPU. We estimate that the GPU version of Matrix eQTL would provide at least tenfold increase in the performance (see Appendix). We will also consider using Matrix eQTL optimization techniques for fast estimation of more complex models, such as multi-SNP models and generalized linear model.

\vskip 0.2in

\textbf{Appendix.}

\vskip 0.1in

\textbf{Matrix Multiplication.} Matrix eQTL gains its efficiency by expressing the most computationally intensive part of the analysis in terms large matrix operations, most importantly matrix multiplication. Naturally, the performance of Matrix eQTL dependents strongly on the performance of the employed basic linear algebra subroutine (BLAS). In Table \ref{tabBlas} we compare performance of matrix multiplication for Matlab and different versions of R for Windows. For the comparison we measured the time required to multiply two 4,096x4,096 matrices with elements set to random values uniformly distributed on [0,1]. The standard installation of R for Windows \citep{Rcite} includes a generic BLAS, not optimized for any particular CPU. The test finished in 90 and 80 seconds for 32 and 64 bit versions of R respectively. There is a faster version of BLAS available for R (32 bit only) called ATLAS \citep{whaley2005minimizing}. The matrix multiplication test on R with C2D version of ATLAS library finished in just 15 seconds. Next we tested Matlab and Revolution R, a commercial version of R available free of charge for academic purposes. They both employ Intel Kernel Math Library (KML), which is optimized for Intel CPUs (as in the test machine) and is able to use multiple CPU cores. The test for both programs finished in 4.3 seconds. Intel KML uses all 4 cores of the test machine and demonstrates about 4 times better performance than R with ATLAS library. About twice better performance can be achieved by switching from double to single precision calculations; we did not use this option in Matrix eQTL to avoid loss of accuracy. Single precision calculation are not available in R. 
The last line in the table shows that NVIDIA GTX 480 GPU (graphics processing unit) can offer ten times better performance than the best algorithm for the CPU (central processing unit) of the test machine. Matlab 2010b has limited build-in support for GPU-based calculations.

The complexity of the direct matrix multiplication for square $n \times n$ matrices is $O(n^3)$. 
More asymptotically efficient methods have been developed with complexity $O(n^{\log_2 7}) \approx O(n^{2.81})$ \citep{strassen1969gaussian} and even $o(n^{2.376})$ \citep{coppersmith1990matrix}. However, in practice, the new methods do not beat the direct one even for relatively large matrices ($n \approx 2000$), and in certain circumstances they may experience numerical instability.
% http://software.intel.com/en-us/forums/showthread.php?t=68045

\begin{table}[h]
\begin{center}
\begin{tabular}{llr@{}ll}
	Package & BLAS & \multicolumn{2}{l}{time (sec)} & comment \\ \hline \hline
	R x32 2.12.1 & build-in &\phantom{....} 90& \\
	R x64 2.12.1 & build-in & 80& \\
	R x32 & Atlas C2D & 15& \\	
	Revolution R 4.0 & Intel KML & 4&.3 \\
	Matlab R2010b & Intel KML & 4&.3 \\ \hline
	Matlab R2010b & Intel KML & 2&.2 & single precision\\ \hline
	Matlab R2010b & GPU CUDA & 0&.25 & GTX 480, single precision\\
\end{tabular}
\caption{Performance of different software in multiplying 4,096x4,096 matrices.
\label{tabBlas}}
\end{center}
\end{table}

%\paragraph{Appendix.}

\textbf{Specifications of the computer and software used for testing:} Brand: Lenovo ThinkStation E20; CPU: Intel Xeon X3430 (2.4 ghz, 4 cores, 38.4 gflop); RAM: 16 GB DDR3; OS: Windows 7; Matlab R2010b; Revoluton R Enterprise 4.0 (64 bit).
% 38.4 http://download.intel.com/support/processors/xeon/sb/xeon3400.pdf

\bibliographystyle{plainnat}
\bibliography{eQTL}

\end{document}